\documentclass[11pt,twocolumn,notitlepage]{article}
\usepackage[utf8]{inputenc}
\usepackage{abstract}
\usepackage{authblk}
\usepackage{booktabs}
\usepackage{longtable}
\usepackage{graphicx}
\graphicspath{/Users/principale/Desktop/articoli/EIODM/LaTeX/Figure}
\usepackage[round]{natbib}

\usepackage{scrpage2}
\ifoot[]{} 
\ofoot[\pagemark]{\pagemark}
\title{\textbf{Environmental Influence on Dental Morphology}}
\author[1]{\textbf{Alessandro Riga\thanks{alessandro.riga2@unibo.it}}}
\author[1]{\textbf{Maria Giovanna Belcastro}}
\author[2]{\textbf{Jacopo Moggi-Cecchi}}

\affil[1]{Department of Biological, Geological and Environmental Sciences, University of Bologna}
\affil[2]{Department of Biology, University of Florence}

\begin{document}
\date{}
\setlength{\textfloatsep}{15 pt}

\twocolumn[
\maketitle

\begin {onecolabstract}
\thispagestyle{empty}
Development is a complex phenomenon where the forming phenotype interacts with genetic and environmental inputs. Teeth are an important model for developmental studies and their development has been thoroughly investigated. However, despite of an extensive literature on the genetics of dental development, no studies have yet focused on the environmental influences on dental morphology. Here we aim to test whether and to what extent the environment plays a role in producing morphological variation in human teeth.
We selected a sample of modern human skulls and used dental enamel hypoplasia as an environmental stress marker to identify two groups with different stress levels, referred to as SG (``stressed'' group)  and NSG (``non-stressed'' group). We collected data on the occurrence and the relative development of 15 morphological traits (5 for each molar) on upper molars using a standard methodology commonly used in dental anthropology (ASU-DAS system) and then we compared the frequencies of the traits in the two groups.
Overall, the results suggest that: (i) stressors like malnutrition and/or systemic diseases have a significant effect on upper molar morphology; (ii) stress generates a developmental response which increases the morphological variability of the SG; (iii) the increase in variability is directional, since individuals belonging to the SG have increased cusp dimensions and number.
These results are consistent with the expectations of the morphodynamic model of dental development \citep{Jernvall:1995xw}.

\end{onecolabstract}]

\saythanks{}

\section*{Introduction}
In developmental studies, teeth represent an excellent model, since once their development is completed, they undergo no further changes in size and shape (except from wear), thus recording the developmental process in their phenotype. Our knowledge on the mechanisms of dental development has increased exponentially over the past 20 years. In particular, it has become evident how dental development, like the development of other organs, is regulated by inductive interactions between epithelial and mesenchymal cells \citep {Ruch:1995uq,Thesleff:1981kx,Thesleff:1995jf}. 
The bulk of the research agenda has focused on the genetic and molecular bases (reviewed in \citealt{Bei:2009fr, Jernvall:2000ee}): detailed knowledge is available today on over 300 genes expressed in mouse dental tissues and their patterns of expression from the beginning of the tooth bud to the end of its development (see for example the Gene expression in tooth~---~WWW database, 1996). The focus on the genetics of dental development has made it possible to make progress in our understanding of morphogenesis and variation in dental morphologies; at the same time, it has overlooked other important aspects, such as the environmental influence on dental development and its contribution to the final phenotype. It is probably because of this bias toward the genetic component of dental development research that is common to find in the literature such statements as ``the development of teeth is under strict genetic control'' \citep{Galluccio:2012sp, Thesleff:1996lt} which, from a developmental biology perspective, greatly underestimate the role of the environmental component. Indeed, ``development is phenotypic change in a responsive (plastic) phenotype due to inputs from the environment and the genome. [\dots] Genomic and environmental factors have equivalent and potentially interchangeable developmental effects, effects that depend as much on the structure of the responding phenotype as they do on the specific inputs themselves'' \citep{West-Eberhard:2003tg}. 

The significant role of environmental factors in determining dental variation was reviewed by \cite{Butler:1983rr}, who acknowledged that ``dental phenotypic differences \emph{between} populations do reflect genetic differences'' (p. 288, Italics ours), but at the same time emphasized that the genes ``that affect the dentition do so only through the mediation of a complex ontogeny in which the environment takes part'' (p. 288).
Different aspects of the environmental influence on tooth development in human populations have been studied. One is the relationship between malnutrition and tooth eruption, where malnourished children reveal a delay in the eruption of their teeth \citep{Boas:1927mz, Enwonwu:1973ai, Garn:1965la, Gaur:2012fy}.
Another way to study the environmental influence on dental development is to correlate tooth size and environmental stress. \cite{McKee:1990et} studied the correlation between enamel hypoplasia occurrence, a marker of developmental disruption, and dental size reduction: they found an association between dental reduction and the presence of enamel hypoplasia in the individual. \cite{Garn:1971th} showed how maternal health status can influence the tooth size of the deciduous and permanent dentition of the offspring. Other authors demonstrated that environment plays a role in determining tooth size \citep{Kolakowski:1981fx, Perzigian:1984sj, Townsend:1979hc}.
Similar results come from the study of fluctuating asymmetry, which is considered a measure of stress and developmental instability also influenced by environmental variables \citep{Carter:2011wd, Parsons:1990bs}. \cite{Kieser:1992jo}, studying fluctuating odontometric asymmetry, used alcohol consumption by the mother during the gestation period as an environmental variable, and found a high level of asymmetry in children of alcoholic mothers compared to children of non-alcoholic mothers. In another paper, \citep{Kieser:1997yo} fluctuating asymmetry in children was considered in relation to maternal obesity and smoking; again asymmetry was higher in children of obese and smoking mothers.
The study of monozygotic twin pairs is another research approach to address how differences in the presence, size, and shape of teeth are linked to environmental influence. ``[\dots] Supernumerary tooth formation is influenced not only by genetic factors but also by environmental and epigenetic influences'' (\citealt{Townsend:2012kq}, p. 6). For example, \cite{Townsend:2005vo} found monozygotic twins to be discordant in the number and position of supernumerary teeth. Other studies on monozygotic twins focused on the heritability of discrete traits; in the first half of the last century, the prevailing view was that morphological traits on tooth crowns were under strong hereditary control (\citealt{Scott:1997gs}, p.161). For example, in the case of Carabelli's trait, the most studied trait on upper molar crowns, starting from the 1950s this view began to shift from a simple single-locus model to a more complex one, invoking multi-loci genetic control or environmental effects. As studies on twins increased in number, it became evident how trait expression often differs between identical twins (\citealt{Scott:1997gs}, p.162). 
Lastly, the effects of pathological conditions on dental development \citep{Atar:2010fj} have been described. Several pathological conditions (\emph{e.g.} cystic fibrosis, AIDS, leukemia) cause a number of enamel and dentine defects, acting during tooth development, in the pre-natal and post-natal periods.

All these studies discussed above have some limitations when considered from an evolutionary developmental biology perspective. 
First of all, the majority of them concentrates on metrical traits; when discrete traits are considered, the focus is on characters of the whole dentition, such as congenital absence and supernumerary teeth \citep{Townsend:2005vo}, or on pathological defects \citep{Atar:2010fj}. In studies on environmental influence, normal individual variation which describes the morphology of a single tooth is usually not taken into account.
Secondly, in evolutionary biology, the relevant variation is that occurring at the population level, since morphological differences among taxa often stem from evolutionary processes taking place at the population level \citep{Mayr:1982yb, Jernvall:2000kn}. Few studies have taken into account environmental influences on teeth at the population level: among them, some were on populations who migrated into new environments \citep{Lasker:1945zv, Scott:1992qz}, but they failed to demonstrate plasticity in dental traits, probably because of the kind of stressor tested \citep{Scott:1997gs}.
Thirdly, even when some degree of plasticity is recognized (\emph{e.g.} differences in trait expression between MZ twins), these studies lack a developmental explanation of the mechanism through which this plasticity is produced. This difficulty is particularly evident in heritability studies, where the issue is trying to distinguish between genetic and environmental variance, as if one genotype were bound to always produce the same phenotype and, as such environmental variance is a mere noise altering the correct correspondence between genotype and phenotype. In this framework, any attempt to explain the developmental process becomes sterile. On the other hand, if we accept that variability on which evolutionary factors act is the product of development  \citep{Salazar-Ciudad:2003jh} and that development is the result of the interactions between genetic and environmental inputs and the developing phenotype \citep{West-Eberhard:2003tg}, the mechanism involved in the shaping of variability becomes central to our investigation.

In recent years, thanks to remarkable advances in the field of molecular biology, it has been possible to begin to understand the molecular basis of this mechanism and how the developmental process leads to the final phenotype. Studies on experimental animals have provided information on genes expressed in different tissues during tooth development; this research, together with a broad scale analysis of morphological variability in mammalian dentition, has led to the proposal of a specific model of development for multicuspidate teeth. The developmental unit for cusp development is the primary enamel knot \citep{Jernvall:2000ee}, ``a cluster of cells in the central part of dental epithelium, facing the dental mesenchyme'' \citep{Butler:1956lq, Vaahtokari:1996qm}. The primary enamel knot acts like an embryonic signaling center and expresses the same genes as involved in the organogenesis of other ectodermal organs such as feathers, nails, and glands \citep{Bei:2009fr, Jernvall:2000jw, Jernvall:2000ee}; the majority of these genes belong to one of the signaling pathways of four gene families: BMP, FGF, SHH, and WNT. Furthermore, the dimensions of the primary enamel knot determine the position and dimensions of the secondary enamel knots \citep{Jernvall:2000ee, Jernvall:2000jw}, from which the cusps originate. Lastly, secondary enamel knots express the same genes as expressed in the primary enamel knot, \emph{i.e.}, in the gene network there are no specific signals for each cusp \citep{Jernvall:2000ee}. 
Starting from these clues, \cite{Jernvall:1995xw} developed a model for secondary enamel knots initiation: a new enamel knot initiates when there are enough epithelial cells available to form them; the number of epithelial cells is regulated by the expression of FGFs in extant enamel knots. The formation of cusp patterns is therefore a cascade of events beginning with the formation of the primary enamel knot: ``the relative sizes, numbers, and heights of the cusps [in seal molars] are symmetrically correlated, anterior and posterior, to a major central cusp, as if the development of these additional cusps depends on rate of growth relative to a threshold in time, beyond which further development does not occur'' (\citealt{Weiss:1998lf}, p. 390; \citealt {Jernvall:2000kn}). The dimension of the primary enamel knot regulates the formation of subsequent secondary enamel knots and the dimensions of those regulate the initiation of other enamel knots (and cusps) later in the developmental cascade. 
Tooth morphogenesis is therefore a self-organizing system or a Bateson-Turing Process \citep{Weiss:1998lf}, where ``once a set of differentiation factors is established [\dots] further exogenous signaling is no longer needed'' (\citealt{Weiss:1998lf}, p. 376). When the first enamel knot appears, the tooth primordium has all the information it needs to complete its development. This model for tooth development is usually referred to as the cascade mode for tooth morphogenesis, or the morphodynamic model. 
Today fairly accurate computational models for tooth development are available, taking into account both the genetic network and the cellular interactions \citep{Osborn:1993px, Osborn:2008cr, Salazar-Ciudad:2002oy, Salazar-Ciudad:2010xh}. Starting from the cascade model, these attempt to reproduce extant variability in tooth diversity. For example, \cite{Salazar-Ciudad:2010xh} developed a model accounting for the morphological diversity in seal molars. Following the suggestions of the cascade model, this diversity is obtained by acting on very few initial inputs representing genetic information: one inhibitor and one activator for the differentiation of enamel knots, and one parameter representing growth factors, upregulated in enamel knots, regulating cell proliferation or differentiation.

With this background, the aim of this paper is to test whether environmental inputs can affect dental morphology acting on the developmental cascade. \cite{Scott:1997gs} suggested that stressors such as undernutrition and infectious diseases, already known to disrupt dental development leaving marks on teeth, could affect dental morphology as well. The occurrence of dental enamel hypoplasia is a good marker for these kind of stressors, since it is known to be linked to physiological stress, such as malnutrition or systemic diseases, during tooth development \citep{Guatelli-Steinberg:1999rw, Ten-Cate:1998jy, Skinner:1992ix}. The response of the developing tooth to such stress is a slowing down of its development and the visible result on the tooth surface is an enamel hypoplastic defect.
\cite{Jernvall:2000kn} suggested that studies on variability at the population level can be useful in testing hypotheses in development biology: our hypothesis is that, in a homogeneous population, individuals with severe hypoplasia, that is to say, individuals whose development was probably disrupted by an environmental stress, are more variable in their molar morphology than individuals with no hypoplasia. The expectation we want to test is that a developmental disruption, due to an environmental stressor, occurring in a population increases variability in molar morphology. In a single organism, a disturbance during the morphogenetic cascade should cause subtle ``errors'' in the morphological output. These changes will be more evident as the ``errors'' accumulate in the developmental cascade, that is to say, in later developing cusps.  If we look at the whole population, where disruption has occurred at different times and with different intensity in each individual, the logical expectation is an increase in morphological variability, particularly evident in later developing cusps. 

\section*{Materials and Methods}
\begin{figure}[]
\centering
\label{fig:1}
\includegraphics[width= \columnwidth]{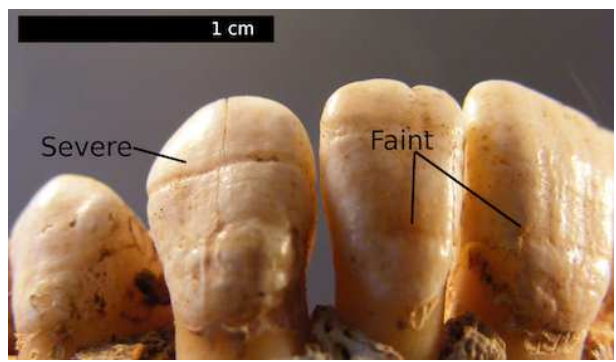}
\caption{\footnotesize{Linear enamel hypoplasia on the maxillary dentition: on the basis of the breadth and depth of the bands, severe and faint grades are distinguished (following \citealt{Buikstra:1994gf}).}}
\end{figure}

The specimens analyzed derive from the ``Fabio Frassetto collection'' housed in the Museum of Anthropology at the University of Bologna. The largest part of this collection is made up of over 600 skeletons recovered from the cemeteries of the town of Sassari (Sardinia, Italy) in the early $20_{th}$ century. For the vast majority of the skeletons, individual information is available (name, sex, date of birth, date of death, job, and cause of death). This sample can thus be considered very homogenous in terms of genetic structure of the population and physical environment.
Selection criteria required the presence of the maxillary dentition, with at least two molar teeth from the same side. Specimens where attrition precluded analysis of cusp morphology were excluded. 
The occurrence of enamel hypoplasia was employed here as marker of environmental stress (\emph{sensu lato}). Enamel hypoplasia is known to be linked to physiological stress, such as malnutrition or systemic diseases, during dental development \citep{Guatelli-Steinberg:1999rw, Ten-Cate:1998jy, Skinner:1992ix}. The presence of linear enamel hypoplasias was recorded following \cite{Buikstra:1994gf}, taking into account the severity with a two-grade scale on the basis of the breadth and depth of the hypoplastic bands (fig. 1\ref{fig:1}). 
Two groups were then identified, with different stress levels: the first group included individuals with no signs of enamel hypoplasia in any of the present teeth; the second group included individuals with at least one tooth with one severe or three faint hypoplastic events.
Only individuals with a clear pattern of presence/absence of hypoplasia have been included. The final sample included 75 individuals (34 females, 41 males). Group 1 (no hypoplasia) included 30 individuals (13 females, 17 males); group 2 (severe or diffuse hypoplasia) included 45 individuals (21 females, 24 males). For the sake of convenience, from now on group 1 and group 2 will be referred to as ``non-stressed'' (NSG) and ``stressed'' (SG), respectively.
For each individual of each group, the variability in the expression of maxillary molar cusps was observed. Expression of the relative development of the cusps was recorded in a semi-quantitative way using the ASU-DAS standard \citep{Turner:1991yu}. Discrete traits considered were: metacone, hypocone, metaconule, Carabelli's trait, and parastyle. Each trait was scored on a scale from 0 to 5 (metacone, hypocone and metaconule), 0 to 6 (parastyle), or 0 to 7 (Carabelli's trait), where 0 is the absence of the trait.
A two-sample Wilcoxon test was used to compare each discrete variable between the two groups. $T^2$ Hotelling test and two-sample MANOVA (Pillai test) were used for multivariate analysis. The $p-values$ were corrected for repeated tests with Bonferroni's method. All the statistical analyses were carried out using the software R \citep{R-Development-Core-Team:2010fb}.

\section*{Results}
The analysis was carried out for each group (NSG and SG) with sex pooled; any further subdivision would have reduced sample sizes and affected reliability of the statistical analysis. Table 1 shows, for each molar and for each morphological trait (cusp), the comparison between NSG and SG in the frequencies of expression of each category. 
A general trend appears in most cases: in the SG, the variability of expression of the different grades (the frequency of each category) is greater than in the NSG and the higher grades of expression are more frequent. That is, in 10 of the 15 cusps considered (metacone on $M^1$ and $M^2$, hypocone and metaconule on $M^1$ and $M^3$, Carabelli's trait and parastyle on $M^2$ and $M^3$), the mode of the distribution is constant, while the mean increases. This condition is suggestive of a directional increase in cusp expression variability in the SG. 
These differences in the distributions are significant (two-samples Wilcoxon test, $p-value < 0.05$) for the following cusps: Carabelli's trait on each of the three molars; metacone on the first, hypocone on the second and metaconule on the third molar. When the $p-values$ are corrected using Bonferroni's method, the values under the threshold of significance are those of hypocone on the second molar and Carabelli's trait on the first and second molars. The variability for parastyle is too low to make sense of it using any statistical test, but it is worth noting that the only two parastyles observed in the sample belong to two individuals of the SG.
As a further step in the study it was deemed suitable to perform a multivariate analysis in order to take into account at the same
\onecolumn

\begin{longtable} {ccccccc}

\caption{\footnotesize{$p-values$ for the Wilcoxon test applied to compare the distributions of cusp development between the non-stressed (NSG) and stressed (SG) groups. The threshold of significance was set at 0.05; the values that maintain their significance after Bonferroni's correction ($p-value < 0.0033$) are in bold.}}\\
\toprule
\toprule
{}&\multicolumn{2}{c}{$M^1$}&\multicolumn{2}{c}{$M^2$}&\multicolumn{2}{c}{$M^3$}\\
\endfirsthead
\multicolumn{7}{l}{\footnotesize\itshape\tablename~\thetable:
continues from previous page} \\
\toprule
\midrule
\endhead
\midrule
\multicolumn{7}{r}{\footnotesize\itshape\tablename~\thetable:
continued on the next page\dots} \\
\endfoot
\bottomrule
\multicolumn{7}{r}{}
\endlastfoot
{Metacone}&\multicolumn{2}{c}{($p=0.037$)***}&\multicolumn{2}{c}{($p=0.431$)ns}&\multicolumn{2}{c}{($p=0.289$)ns}\\\hline
\midrule
{} & {NSG} & {SG} & {NSG} & {SG} & {NSG} & {SG}\\
0&	0&	0&	0&	0&	0&	0\\
1&	0&	0&	0&	0&	0&	0\\
2&	0&	0&	0&	1&	1&	0\\
3&	0&	0&	4&	8&	10&	18\\
4&	12&	9&	35&	39&	15&	16\\
5&	31&	64&	12&	26&	0&	9\\
\bottomrule
{}&\multicolumn{2}{c}{$M^1$}&\multicolumn{2}{c}{$M^2$}&\multicolumn{2}{c}{$M^3$}\\
{Hypocone}&\multicolumn{2}{c}{($p=0.302$)ns}&\multicolumn{2}{c}{\textbf{(p=0.002)***}}&\multicolumn{2}{c}{($p=0.408$)ns}\\\hline
\midrule
{} & {NSG} & {SG} & {NSG} & {SG} & {NSG} & {SG}\\
0&	0&	0&	16&	5&	11&	16\\
1&	0&	0&	8&	9&	4&	6\\
2&	0&	0&	3&	11&	5&	3\\
3&	4&	4&	10&	18&	6&	14\\
4&	15&	22&	14&	17&	1&	4\\
5&	23&	46&	0&	11&	0&	0\\
\bottomrule
{}&\multicolumn{2}{c}{$M^1$}&\multicolumn{2}{c}{$M^2$}&\multicolumn{2}{c}{$M^3$}\\
{Metaconule}&\multicolumn{2}{c}{($p=0.362$)ns}&\multicolumn{2}{c}{($p=0.580$)ns}&\multicolumn{2}{c}{($p=0.008$)***}\\\hline
\midrule
{} & {NSG} & {SG} & {NSG} & {SG} & {NSG} & {SG}\\
0&	40&	64&	46&	68&	26&	33\\
1&	0&	5&	3&	4&	0&	5\\
2&	1&	0&	0&	0&	0&	1\\
3&	1&	1&	1&	0&	0&	2\\
4&	0&	1&	0&	0&	0&	0\\
5&	0&	0&	0&	0&	0&	2\\
\bottomrule
\multicolumn{7}{c}{}\\
\multicolumn{7}{c}{}\\
\multicolumn{7}{c}{}\\
\multicolumn{7}{c}{}\\
{}&\multicolumn{2}{c}{$M^1$}&\multicolumn{2}{c}{$M^2$}&\multicolumn{2}{c}{$M^3$}\\
{Carabelli's trait}&\multicolumn{2}{c}{\textbf{(p=0.000)***}}&\multicolumn{2}{c}{\textbf{(p=0.001)***}}&\multicolumn{2}{c}{($p=0.031$)***}\\\hline

\midrule
{} & {NSG} & {SG} & {NSG} & {SG} & {NSG} & {SG}\\
0&	25&	9&	48&	50&	26&	31\\
1&	6&	15&	2&	14&	1&	2\\
2&	7&	8&	1&	4&	0&	5\\
3&	3&	9&	0&	2&	0&	0\\
4&	2&	12&	0&	1&	0&	1\\
5&	0&	13&	0&	0&	0&	1\\
6&	0&	1&	0&	0&	0&	0\\
7&	0&	3&	0&	0&	0&	0\\
\bottomrule
{}&\multicolumn{2}{c}{$M^1$}&\multicolumn{2}{c}{$M^2$}&\multicolumn{2}{c}{$M^3$}\\
{Parastyle}&\multicolumn{2}{c}{na}&\multicolumn{2}{c}{\textbf{(p=0.416)ns}}&\multicolumn{2}{c}{($p=0.440$)ns}\\\hline
\midrule
{} & {NSG} & {SG} & {NSG} & {SG} & {NSG} & {SG}\\
0&	43&	73&	51&	73&	27&	41\\
1&	0&	0&	0&	0&	0&	0\\
2&	0&	0&	0&	0&	0&	0\\
3&	0&	0&	0&	0&	0&	1\\
4&	0&	0&	0&	1&	0&	0\\
5&	0&	0&	0&	0&	0&	0\\
6&	0&	0&	0&	0&	0&	0\\
\bottomrule

\end{longtable}

\twocolumn
time the grade of development of every cusp in each tooth or in each single individual and compare it between the two groups. 
This was done since the scope of the work is to address the overall variability in cusp expression of the molar teeth, rather than the relative development of each single cusp.
A principal component analysis considering, for each individual and each group, all three molar teeth and the grade of expression of each cusp (except for the parastyle because of its very low frequency in the sample) was carried out.
\begin{figure*}[p]
\centering
\label{fig:2}
\includegraphics[width= 2\columnwidth]{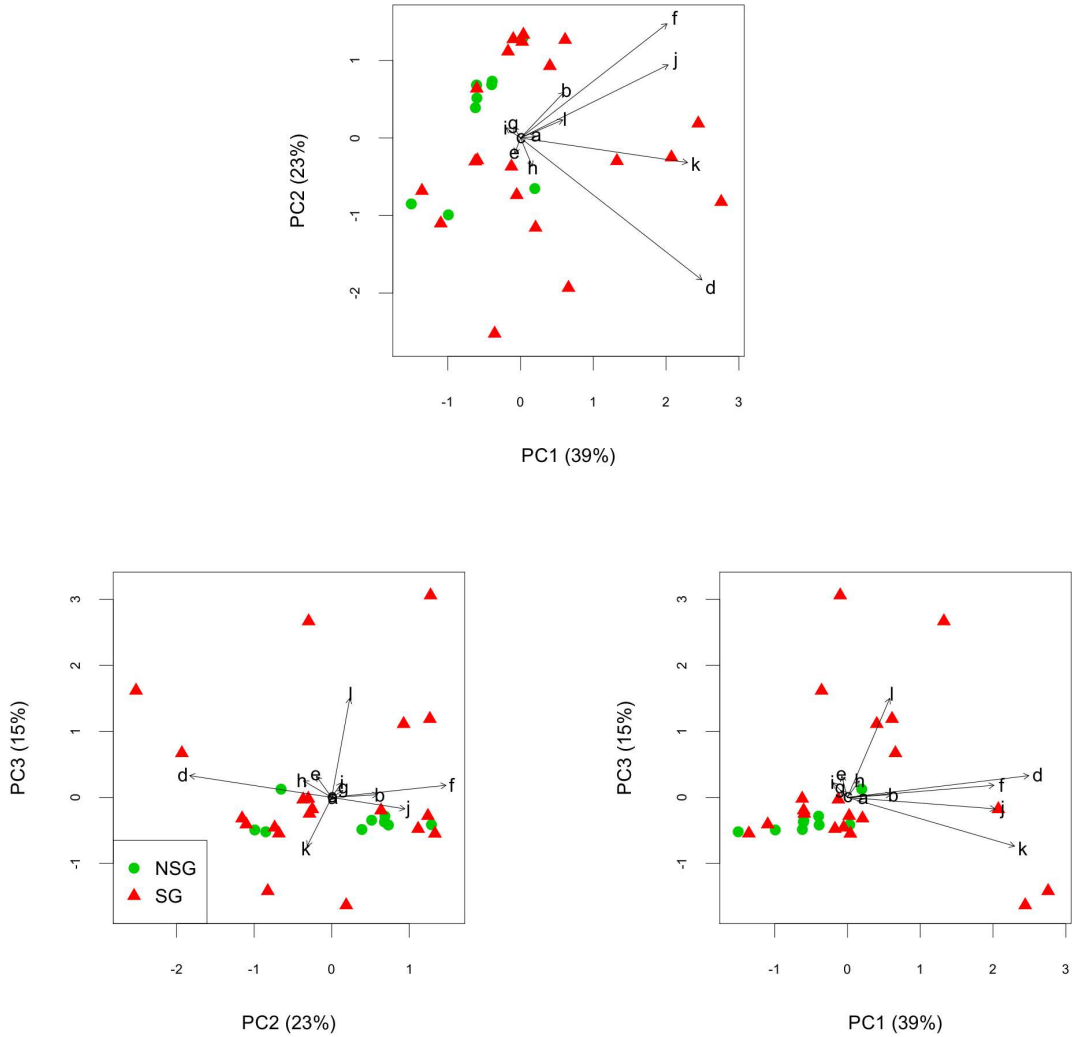}
\caption{\footnotesize{Plots of the first three PCs for all the traits in upper molars (parastyle was excluded because of the low frequency in the sample). The arrows represent the weight of each variable on the PCs (a=metacone on $M^1$; b=hypocone on $M^1$; c=metaconule on $M^1$; d=Carabelli's trait on $M^1$; e= metacone on $M^2$; f= hypocone on $M^2$; g= metaconule on $M^2$; h= Carabelli's trait on $M^2$; i= metacone on $M^3$; j= hypocone on $M^3$; k= metaconule on $M^3$; l= Carabelli's trait on $M^3$).}}
\end{figure*}
 Figure 2\ref{fig:2} shows scatterplots of the first three principal components in pairs, where individuals of the NSG and SG are indicated in different colors. The first three principal components explain the 77\% of the variability of the dataset ($PC1=39\%$, $PC2=23\%$, $PC3=15\%$). In Figure 2 the weight of each variable on principal components is also shown (black arrows). Variation along PC1 is linked to Carabelli's trait on $M^1$, hypocone on $M^2$ and $M^3$, and metaconule on $M^3$; variation on PC2 is mostly explained by Carabelli's trait on $M^1$ and hypocone on $M^2$ and $M^3$; lastly, metaconule and Carabelli's trait on $M^3$ explain most of the variation along PC3. The same analysis was also performed separately for each molar tooth.
A similar pattern in the spatial distribution of the individuals belonging to the two groups emerges in both types of analysis. Individuals from the NSG tend to cluster together in a relatively delimited area of the plot, whereas those from the SG are more broadly distributed on the Cartesian plane, partially overlapping the NSG. 
PC1 values for the NSG vary between $-3.19$ and $0.42$, while in the SG between $-2.88$ and $5.86$; PC2 values between $-1.62$ and $2.10$ for the NSG and between $-4.12$ and $2.17$ for the SG; PC3 values between $-0.70$ and $0.17$ (NSG) and $-2.18$ and $4.09$ (SG). This indicates a higher variability in the SG as compared to the NSG, more easily detectable on PC1, which best distinguishes the two groups, followed by PC3.
Also, individuals in the NSG tend to fall into the margin of the distribution, suggesting that the enhanced variability of the SG is not isotropic but directional. Once again, the direction toward which variability increases is the direction of more developed cusps (indicated by the direction of the arrows in Figure 2).
In order to test whether the differences in the observed distributions were statistically significant, a Hotelling's $T^2$ test was employed; first with the three molars pooled, and then for each molar separately. In the first case, parastyle and metaconule had to be excluded from the analysis, because of their overall very low frequencies and the reduced number of cases due to the omission of strings with NAs data. On the other hand, in the analysis of each molar tooth (where reduced variables for each case made it possible to omit fewer strings), it was possible to include metaconule, but not parastyle, in the variables considered. 
\begin{table}[tb2]
\centering
\caption{\footnotesize{Multivariate comparison (Hotelling $T^2$ test) between SG and NSG, for the molars pooled ($M^1$+$M^2$+$M^3$) and for each molar separately. Significance threshold is set to $p=0.05$ ($p=0.0125$ after Bonferroni's correction).}}
\label{tab:2}
\begin{tabular*}{\columnwidth}{@{\extracolsep{2cm} }rc}
\toprule
{Tooth}&{$p-value$}\\
{}&{Hotelling $T^2$}\\
\midrule
{$M^1+M^2+M^3$} & 0.00059\\
{$M^1$}&0.00000\\
{$M^2$}&0.00004\\
{$M^3$}&0.00178\\
\bottomrule
\end{tabular*}
\end{table}
In both analyses, differences between NSG and SG are statistically significant (Tab. \ref{tab:2}), even after Bonferroni's correction for repeated tests.

\section*{Discussion}
The different degree of expression of some dental morphological traits in the two groups of a modern human population (SG and NSG) seems to suggest the existence of a link between enamel hypoplasia occurrence and upper molar teeth morphology, in that both are the result of the same cause: a developmental disruption, likely due to a non-genetic disturbance such as malnutrition or a systemic disease (among the most common causes of enamel hypoplasia). In other words, an environmental disturbance affecting dental development has an influence on the final phenotype.
Two aspects of the results are worthy of note: first, as we hypothesized, a developmental disruption resulted in an increased variability in dental morphology; secondly, the increase in variability is directional, toward an expression of larger cusps. These two aspects will now be discussed in detail.

\subsection*{Variability increase in the stressed group}
The first effect of the developmental disruption on the expression of upper molars non metric traits is an increase of variation in the SG.
The increase of variability in the expression of morphological traits under a stressful condition has usually been attributed to two different factors. The first is related to the existence of a hidden genetic variability that is canalized by a buffering mechanism in normal development \citep{Waddington:1959li}; if an environmental stressor exceeds a threshold, the buffering mechanism breaks down and new genetic variability arises \citep{Rutherford:2000zh}. However, there are few cases in which this buffering mechanism has been discovered \citep{Rutherford:1998sh, Sollars:2003lc} and there are no clues indicating that a similar genetic buffering mechanism may be active during dental development.
The other possibility discussed in the literature is that new variability arises from the developmental mechanism itself \citep{Salazar-Ciudad:2007dl}. As mentioned in the introduction, the current developmental model proposed for the molars has, as a logical consequence, an increase of variability in a developmentally disrupted population. Indeed, molar tooth development is a morphodynamic process in which morphogenesis and induction mechanisms are interdependent \citep{Jernvall:1995xw, Salazar-Ciudad:2003jh}. In morphodynamic systems a small change in the inputs can be amplified during the morphodynamic process (developmental cascade) and have unexpected results on the final morphology, because the relationship between the parameters' change and morphology is not linear, but depends on the developmental network, on the phase of morphogenesis, and on all the other inputs involved in tooth development. Therefore, the effect we observed in our sample on the increase of morphological variability in the SG is compatible with this current view of tooth development. 
Development is an epigenetic phenomenon; therefore, if we wish to understand the mechanism responsible for our results, we must look at the epigenetic level, where dental development is regulated by cellular and genetic interactions \citep{Salazar-Ciudad:2010xh}. Both these factors (cellular parameters and gene expression) can be affected by environmental inputs. 
As for the first factor, environmental stress is known to produce a downregulation of cellular metabolism \citep{Hand:1996ec}. The effects of this downregulation can be multiple; for example during hypometabolic states (such as starvation or lack of nutrients), downregulation can affect both energy production and energy consumption, and can inhibit macromolecule synthesis and turnover. In a similar situation it is likely that other parameters important for tooth development may also change, such as, for example, epithelial and mesenchymal growth and differentiation rates. In this regard, some authors \citep{Johnston:2002pt, Katso:2001ay, Kozma:2002zh, Schmelze:2000iw} have suggested that cell growth is affected by environmental and developmental conditions and that, in some cases, it can change in response to nutrient levels. \cite{Kim:2002uf} identified Raptor, a protein sensitive to nutrient levels participating in the regulation of cell size.
As for the importance of gene expression, there is a vast literature on how environmental stressors can alter gene expression, even just considering nutritional stressors. In bees, for example, queens and sterile workers differentiate from genetically identical larvae thanks to a different nutrition that probably influences the expression of some genes through DNA methylation \citep{Kucharski:2008ta}. Also, in \emph{Drosophila}, gene expression is regulated by starvation \citep{Zinke:2002dt}. In mammals, diet can influence the pattern of methylation and the stable expression of some genes \citep{Jaenisch:2003qf}; nutrition is important during the gestational period and can affect the phenotype of the offspring, for example inducing changes in methylation in rats  \citep{Burdge:2007ul}. Most of the literature on the effects of the environment on gene expression focuses on regulation by changes in DNA methylation, but it is not the only possibility. Transposable elements can also regulate gene expression under stressful situations \citep{Slotkin:2007dd}. For example in mammals, apoptosis regulation is participated by a LTR retrotransposon, \citep{Romanish:2007ri} and retrotransposon activation is known to be stress-induced \citep{Frucht:1991dp, Hampar:1976fe, Hohenadl:1999bv, Ruprecht:2006pz, Sutkowski:2004yi}.
In humans, undernutrition can alter the growth trajectory of babies; also, their morphology and physiology is affected by the nutritional state of the mother \citep{Bateson:2004rt}. \cite{Somel:2008nq} found that differences in diet can produce differences in gene expression. They replicated human and chimpanzee diets in laboratory mice and, after two weeks, they analyzed the differences in gene expression in the liver and in the brain. They found that 4-8\% of genes expressed in the liver changed their expression; considering that only 15\% of the genes expressed in the liver are differently expressed between humans and chimpanzees, these data are impressive.
Thus during dental development, from teeth primordia in intrauterine life up to calcification of the permanent teeth, the environment can act on all the factors listed above. Changes in one or more of those inputs can affect tooth development, producing a wide range of morphological variation. 

\subsection*{Directionality of variability increase}
The other, somehow unexpected, main finding of the analysis is related to the evident directionality in the increase of the variability. This is quite apparent in the PC plots, where individuals belonging to the SG are scattered in the Cartesian plane not randomly with respect to the NSG individuals. In particular, along PC1, NSG individuals tend to fall mostly toward negative values of the first component, whereas SG individuals are uniformly distributed along the axis. Thus variation increases in the direction of more developed cusps (fig. 2), a pattern which is also the most common in the univariate comparisons of each cusp. This result seems to be in contrast with the results of \cite{McKee:1990et}, who reported a correlation between enamel hypoplasia occurrence and dental size reduction. This is not the case, since the ASU-DAS system, used in this study, takes into account the relative development of each trait in the tooth and not the absolute dimensions.
We then considered two options to evaluate the source of the directionality observed:
(i) it is a mathematical artifact due to the presence of a left wall in the distribution of ASU-DAS grades; (ii) it is a true directionality, given by the particular stressor considered, which alters development in a specific direction.

\subsubsection*{Directionality as a mathematical artifact}
\begin{figure}[]
\centering
\label{fig:3}
\includegraphics[width= \columnwidth]{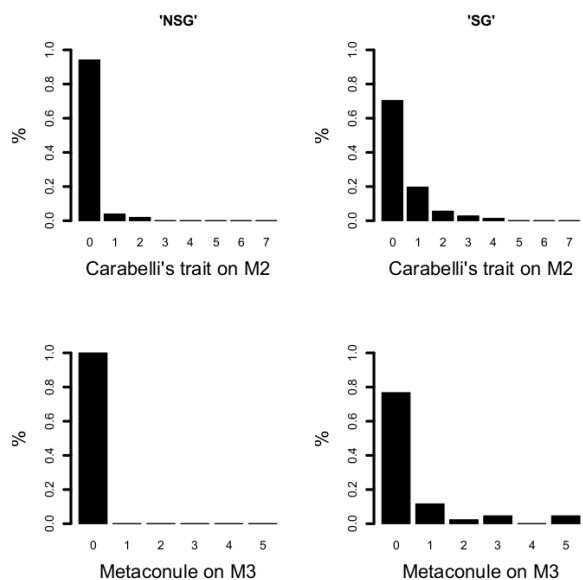}
\caption{\footnotesize{Differences between SG (left column) and NSG (right column) for Carabelli's trait on $M^2$ and metaconule on $M^3$. The directional increase in the mean of the SG could be a mathematical artifact due to a non-directional increase in variability in an asymmetric distribution with a left wall (absence of the trait) and a mode standing next to the wall.}}
\end{figure}

In general terms, when comparing two distributions with different mean values, it is reasonable to consider the distribution with a larger mean value as the effect of some directional factor. However, in some cases, it may happen that a non-directional increase in variability results in a directional change in the distribution, without any directional factor in action \citep{Gould:1996xr}. This is the case for asymmetrical distributions with a left wall, a mode standing constantly near the wall and a right tail of variable length. The increase in the mean may be explained as an increase in variability which, because of the asymmetric distribution, is resolved at all in the right tail.
In our sample (considering only statistically significant comparisons between NSG and SG), Carabelli's trait on $M^2$ and metaconule on $M^3$ seem to fall in this category. The mode of the NSG is 0 (absence of the trait) and variations from this state are absent or very rare; in the SG the mode does not change and more individuals fall into the right tail, which is also longer than in the NSG (fig. 3\ref{fig:3}). The increase in variability is resolved at all in the right tail, because values lower than 0 (the left wall) simply do not exist. 
Therefore, the directionality in changes in morphology in the SG might derive from some non-directional factor acting at the epigenetic level. This could mean that the environmental stressor considered acts on a wide range of epigenetic parameters, without any specificity, randomly with respect to the effects on the developmental cascade.

\subsubsection*{Directionality as real}
\begin{figure}[]
\centering
\label{fig:4}
\includegraphics[width= \columnwidth]{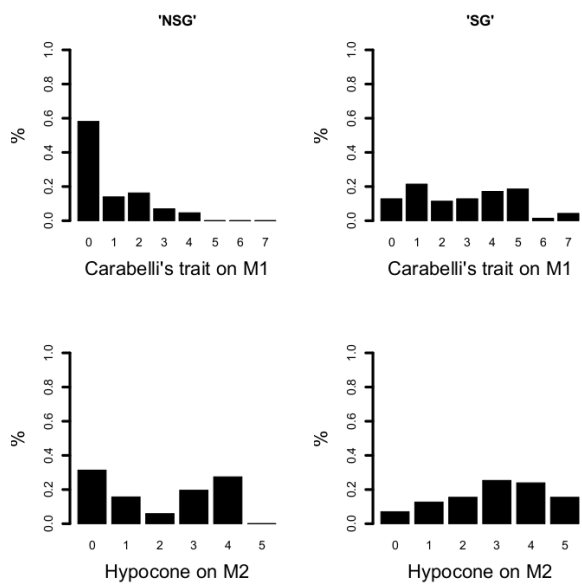}
\caption{\footnotesize{Differences between SG (left column) and NSG (right column) for Carabelli's trait on $M^1$ and hypocone on $M^2$. In this case, the increase in the mean of the SG cannot be due to a mathematical artifact, since the mode changes also; the most probable explanation is the action of a directional factor on the developmental response.}}
\end{figure}

The likelihood of directionality as a mathematical artifact, however, does not fit with other results. Again, considering statistically significant results only, Carabelli's trait on $M^1$ and hypocone on $M^2$ show a different pattern of distribution in the comparison between NSG and SG (\ref{fig:4}fig. 4). In these cusps the mean value increases in the SG but, unlike the former cases, the mode also increases in the SG. In particular for the Carabelli's trait on $M^1$, the whole distribution in the SG appears to be shifted toward higher grades, with a decrease in the ``grade 0'' and an increase in all the other grades.

This evidence seems to support an explanation based on directional factors acting at the epigenetic level. Since dental development is an auto-organizing system, with a few starting inputs and a consequential cascade of events, with no apparent specific signals for different cusps, it is conceivable that the pattern of directional increase in the variability in the SG is produced by the same mechanism for all cusps. For this reason, we cannot rely on an explanation based both on a non-directional and a directional factor. At the moment, a directional factor acting at the epigenetic level seems to us to be the most probable explanation, since it can account for both types of distributions observed, whereas a non-directional factor fails to explain the distributions for Carabelli's trait on $M^1$ and hypocone on $M^2$. 
If we accept this hypothesis, then we might expect that the stressor considered acts not randomly on the variety of parameters affecting dental development. Among these parameters, regulation of signaling during cusp development seems to be the most important source of variation among individuals, while cellular parameters best explain variation along the tooth row \citep{Salazar-Ciudad:2010xh}.
Considering only molecular signaling, cusps pattern is controlled by a game of interactions between signaling molecules regulating epithelial and mesenchymal cell proliferation, such as FGF-4, BMP, and Shh \citep{Jernvall:1994os, Vaahtokari:1996qm}. These interactions regulate the induction of new enamel knots, that is to say they produce the cusp pattern in a forming tooth. BMPs (Bone Morphogenetic Proteins), in particular, are associated with enamel knot apoptosis and its termination, for example inducing inhibitors of cell proliferation like \emph{p21} \citep{Jernvall:1998qq} and \emph{ectodin} \citep{Kassai:2005dw}; studies on \emph{ectodin} expression demonstrated that \emph{ectodin}-deficient mice have extra teeth and extra cusps originated from a larger primary enamel knot \citep{Kassai:2005dw}. Therefore, an environmental stressor could produce the pattern of increased cusp size and number simply inhibiting the production of BMPs, or BMP-induced molecules. Studies on gene expression during dental development are needed to confirm or reject this hypothesis.

As a final remark, we would like to consider the possible evolutionary significance of the kind of plasticity observed in dental development. It has been shown that the evolutionary history of mammalian dental pattern is characterized by repeated convergent or parallel evolution of new cusps such as the hypocone \citep{Hunter:1995kb, Jernvall:2000kn, Jernvall:1996xq}, whereas cusp loss is an uncommon event. That is to say that the developmental response to environmental stress appears to be in the same direction as the evolutionary trend in mammalian dentition. There are several possible explanations. For example, the developmental response could be simply the mark of a developmental constraint biasing variation in a certain direction; a non-isotropic variation due to developmental reasons may result in a directionality imposed to evolution\citep{Gould:1977oz, Gould:2002dk}. Another intriguing possibility is that developmental plasticity in dental development plays an active role in evolution. Developmental plasticity is a topic that has received increasing attention in evolutionary studies \citep{Aubret:2009yq, Chapman:2000eu, Emlen:2007qe, Gibson:1996pb, Price:2003bx, Price:2006ya, Rutherford:1998sh, Sollars:2003lc, Stauffer:2004ec, West-Eberhard:2003tg}. For example, plasticity in certain morphological traits could allow survival in difficult times; phenotypic changes, with the time, may accommodate and lead to a genetic change in the same direction \citep{Pigliucci:2003ct, Pigliucci:2006fi}. Scholars are still debating this topic and there is no consensus \citep{Jong:2005hq, Pigliucci:2006fi}, but it is certainly an interesting avenue for future investigations~---~also with special reference to dental evolution~---~in particular because nutritional stress as a result of environmental pressures is common in many species and may have played a significant role in human and mammalian evolutionary history.

\section*{Conclusions}
The results of our analysis seem to support the hypothesis that an environmental stress can lead to changes in dental morphology. In the group with high levels of enamel hypoplasia, an increase in cusp dimension variability was identified. Further, the increase is not isotropic, but directional: individuals with hypoplasia tend to have relatively larger cusps than individuals without hypoplasia. 
These results open up new possibilities for future research; in particular it would be interesting to inquire into the molecular mechanisms acting at the epigenetic level producing the developmental plasticity in molar morphology, for example testing the environmental influence on the expression of genes for BMPs or BMP-induced inhibitors such as \emph{p21} and \emph{ectodin} (see \citealt{Jernvall:1998qq, Kassai:2005dw}).
Our results confirm, once more, how dental development is a complex phenomenon: it depends on inputs from the gene network and from the environment acting on the developing phenotype. Considering teeth as a strictly genetically determined biological structure, can be misleading in certain kinds of studies.

\section*{Acknowledgements}
We would like to thank Alessio Boattini for his useful advices on {R} and Fabrizio Mafessoni for his assistance in using \LaTeX.
\bibliographystyle{riga_custom}
\nocite{:1996uq} 
\bibpunct{(}{)}{;}{a}{,}{,}
\bibliography{Riga_bibliography}
\end{document}

